\documentclass[sigconf]{acmart}

\usepackage{hyperref}
\usepackage{enumitem}
\usepackage{booktabs} 
\usepackage[table]{xcolor}
\usepackage{colortbl}     
\usepackage{multirow} 
\usepackage{amsmath}
\usepackage{makecell}
\usepackage{caption}
\usepackage{tabularx, array}
\usepackage{xspace}
\usepackage{algorithm}
\usepackage{algorithmic}

\newcommand{\method}{EffiReasonTrans\xspace}
\newcommand{\sftonly}{SFT-Only\xspace}
\newcommand{\rlonly}{RL-Only\xspace}
\newcommand{\dataset}{\method-Data\xspace}
\newcommand{\smallwithmethod}{1.5B\textsubscript{\method}\xspace}
\newcommand{\javatopython}{Java $\rightarrow$ Python\xspace}
\newcommand{\javatocpp}{Java $\rightarrow$ C++\xspace}
\newcommand{\cpptojava}{C++ $\rightarrow$ Java\xspace}
\newcommand{\cpptopython}{C++ $\rightarrow$ Python\xspace}
\newcommand{\pythontocpp}{Python $\rightarrow$ C++\xspace}
\newcommand{\pythontojava}{Python $\rightarrow$ Java\xspace}

\usepackage[most]{tcolorbox}

\newcommand{\boxmargin}{2mm}
\newtcolorbox{mybox}{
    colback=yellow!10!white,
    colframe=gray!50,
    arc = 0pt, outer arc = 5pt,
    boxsep=0pt,
    leftrule=3pt,
    bottomrule=0pt, toprule=0pt, rightrule=0pt,
    left = \boxmargin, right = \boxmargin, top = \boxmargin, bottom = \boxmargin,
    before skip=5pt,
    after skip=5pt
}

\renewcommand\footnotetextcopyrightpermission[1]{}

\begin{document}

\title{{\method}: RL-Optimized Reasoning for Code Translation}

\author{Yanlin Wang}
\email{wangylin36@mail.sysu.edu.cn}
\affiliation{%
  \institution{Sun Yat-sen University}
  \city{Zhuhai}
  \country{China}
}

\author{Rongyi Ou}
\email{oury@mail2.sysu.edu.cn}
\affiliation{%
  \institution{Sun Yat-sen University}
  \city{Zhuhai}
  \country{China}
}

\author{Yanli Wang}
\email{wangyli58@mail2.sysu.edu.cn}
\affiliation{%
  \institution{Sun Yat-sen University}
  \city{Zhuhai}
  \country{China}
}

\author{Mingwei Liu}
\email{liumw26@mail.sysu.edu.cn}
\affiliation{%
  \institution{Sun Yat-sen University}
  \city{Zhuhai}
  \country{China}
}

\author{Jiachi Chen}
\email{chenjch86@mail.sysu.edu.cn}
\affiliation{%
  \institution{Sun Yat-sen University}
  \city{Zhuhai}
  \country{China}
}

\author{Ensheng Shi}
\email{shiensheng@huawei.com}
\affiliation{%
  \institution{Huawei Cloud Computing Technologies Co., Ltd.}
  \city{Beijng}
  \country{China}
}

\author{Xilin Liu}
\email{liuxilin3@huawei.com}
\affiliation{%
  \institution{Huawei Cloud Computing Technologies Co., Ltd.}
  \country{China}
}

\author{Yuchi Ma}
\email{mayuchi1@huawei.com}
\affiliation{%
  \institution{Huawei Cloud Computing Technologies Co., Ltd.}
  \country{China}
}

\author{Zibin Zheng}
\email{zhzibin@mail.sysu.edu.cn}
\affiliation{%
  \institution{Sun Yat-sen University}
  \city{Guangzhou}
  \country{China}
}

\renewcommand{\shortauthors}{Trovato et al.}

\begin{abstract}
    Code translation is a crucial task in software development and maintenance. While recent advancements in Large Language Models (LLMs) have improved automated code translation accuracy, these gains often come at the cost of increased inference latency—hindering real-world development workflows that involve human-in-the-loop inspection. To address this trade-off, we propose \method, a training framework designed to improve translation accuracy while balancing inference latency. We first construct a high-quality reasoning-augmented dataset by prompting a stronger language model DeepSeek-R1 to generate intermediate reasoning and target translations. Each (source code, reasoning, target code) triplet undergoes automated syntax and functionality checks to ensure reliability. Based on this dataset, we employ a two-stage training strategy: supervised fine-tuning on reasoning-augmented samples, followed by reinforcement learning to further enhance accuracy, which also helps balance inference latency. We evaluate \method\ on six translation pairs. Experimental results show that \method\ consistently improves translation accuracy (up to +49.2\% CA and +27.8\% CodeBLEU compared to the base model), while reducing the number of generated tokens (up to -19.3\%) and lowering inference latency in most cases (up to -29.0\%). Ablation studies further confirm the complementary benefits of the two-stage training framework. Additionally, \method\ shows improvements of translation accuracy when integrated into agent-based frameworks. Our code and data are available at \url{https://github.com/DeepSoftwareAnalytics/EffiReasonTrans}.   
\end{abstract}

\maketitle

\section{Introduction}
Code translation is a crucial task in software development and maintenance, involving converting source code from one programming language into another to suit different application scenarios~\cite{zhong2010mining,lachaux2020unsupervised,yang2024exploring,liu2023syntax,chen2018tree, nguyen2015divide, liu2024hmcodetrans}. Recent studies reveal that automated code translation techniques based on Large Language Models (LLMs) are promising~\cite{jana2024cotran,yang2024exploring,tao2024unraveling,yuan2024transagent, liu2024hmcodetrans,ibrahimzada2024alphatrans, eniser2024towards, zhang2025scalable, yin2024rectifier, pan2024lost,macedo2024intertrans, bhattarai2024enhancing}. For example, UniTrans\cite{yang2024exploring} enhances code translation accuracy by generating test cases and iteratively repairing errors using LLMs, while hmCodeTrans\cite{liu2024hmcodetrans} leverages human-machine collaboration to improve accuracy. While these approaches have improved translation accuracy, challenges remain in handling complex translation scenarios. To address such challenges, explicit reasoning methods like Chain-of-Thought (CoT) prompting have been explored, which has been shown to improve performance on a range of arithmetic, commonsense, and symbolic reasoning tasks~\cite{wei2022chain,ho2022large,chung2024scaling,zhou2022least}. Recent studies have incorporated them into the code translation task to improve performance~\cite{tao2024unraveling,yuan2024transagent}. More recently, models like DeepSeek-R1 have been designed to generate internal reasoning processes without prompting~\cite{guo2025deepseek}, reflecting continued interest in utilizing the powerful reasoning capabilities of LLMs. 
    
However, the enhanced performance comes at the cost of longer inference chains, {resulting in substantial increases in inference latency}. Our preliminary experiments show that while reasoning model (i.e., DeepSeek-R1) improves translation accuracy from 86\% to 92\% compared to non-reasoning model (i.e., DeepSeek-V3) on Unitrans-Dataset~\cite{yang2024exploring}, there is a 540\% increase in inference latency. In interactive development scenarios, developers typically expect rapid and reliable feedback, making inference speed a critical factor alongside translation accuracy. This is especially relevant in human-in-the-loop pipelines such as hmCodeTrans~\cite{liu2024hmcodetrans}, where developers participate in reviewing and correcting model-generated translations. In large-scale industrial systems, this interaction is often indispensable, and {high inference latency can significantly hinder the development workflow}. Therefore, we aim to answer this question: \textbf{how to balance accuracy and efficiency while harnessing LLMs' powerful reasoning capabilities? }

In this paper, we propose \textbf{\method}, a training framework that integrates reasoning-augmented data synthesis with a two-stage training process to improve code translation accuracy while optimizing inference latency. Specifically, \method consists of three key stages: data synthesis, supervised fine-tuning, and reinforcement learning. \textcircled{1} \method first synthesizes high-quality (source code, reasoning, target code) triplets by leveraging a more capable language model (DeepSeek-R1~\cite{guo2025deepseek}), filtered via automated syntax validation and functional testing to filter out incorrect or incomplete samples. This process produces \textbf{\dataset}, a reliable reasoning-augmented code translation corpus that capture the semantic and logical migration from source to target language. 
\textcircled{2} Next, \method employs supervised fine-tuning followed by reinforcement learning (using the GRPO algorithm~\cite{shao2024deepseekmath}), guided by a custom reward strategy with dual objectives: (1) execution correctness (test case pass rate) and (2) output conciseness (length tolerance), jointly reducing latency without sacrificing accuracy. 

To evaluate the effectiveness of \method, we conduct extensive experiments on 6 translation pairs among Python, Java, and C++ and obtain the following findings. 
\textcircled{1} Experiments demonstrate that \method achieves superior accuracy while optimizing latency. For instance, on \javatopython, it improves CA by 27.4\%, APR by 23.1\%, and CodeBLEU by 10.1\%, while reducing latency by 29.0\%. 
\textcircled{2} Ablation studies reveal that both supervised fine-tuning and reinforcement learning contribute to these gains, with RL further boosting CA by up to 34.0\% and reducing latency by 25.4\%. 
\textcircled{3} Notably, \method maintains strong performance under limited model capacity and benefits from multilingual training data, showcasing its generalizability. 
\textcircled{4} Finally, when integrated into agent-based frameworks, \method preserves its accuracy improvements in end-to-end pipelines.

Our main contributions in this work include:
\begin{itemize}[noitemsep, topsep=3pt]
    \item We propose \method, a two-stage training framework that integrates reasoning-augmented data synthesis with supervised fine-tuning and reinforcement learning, aiming to improve the accuracy-efficiency trade-off in code translation.
    \item We design a task-driven data synthesis method that leverages a stronger language model to generate reasoning augmented code translation triplets, coupled with automated syntax and functional tests to ensure data quality. The resulting high-quality dataset is \dataset.
    \item We design a dual-objective reward strategy for code translation that considers both execution correctness and output conciseness.
    \item We conduct extensive experiments to show that \method effectively improves translation accuracy while reducing latency. 
\end{itemize}

\section{Related Work}
\subsection{Automated Code Translation}
    In recent years, large language models (LLMs) have developed rapidly, with increasing application to tasks such as code translation~\cite{codetrans1, yang2024exploring, ibrahimzada2024alphatrans, liu2024hmcodetrans, yuan2024transagent,codetrans2, codetrans3,codetrans4}, code generation~\cite{codegen1,codegen2,codegen3,codegen4,codegen5,codegen6,codegen7,codegen8,codegen9,codegen10,codegen11,codegen12,codegen13,codegen14,codegen15,codegen16,codegen17}, code summarization~\cite{codesum1,codesum2,codesum3,codesum4}, code search~\cite{codesearch1,codesearch2,codesearch3,codesearch4,codesearch5,codesearch6}, issue resolution~\cite{issue1,issue2,issue3,issue4}, vulnerability detection and repair~\cite{vulner1,vulner2,vulner3,vulner4,vulner5,vulner6}, etc. Among these, automated code translation refers to the task of converting source code from one programming language to another, aiming to support software reuse, cross-platform compatibility, and long-term maintainability  ~\cite{zhong2010mining,lachaux2020unsupervised,yang2024exploring,liu2023syntax,chen2018tree, nguyen2015divide, liu2024hmcodetrans}. Early approaches are dominated by rule-based methods, which relied on handcrafted grammars and language-specific transformation rules~\cite{c2rust,cxgo}. These methods often suffered from low readability and correctness, especially in the face of modern programming paradigms and dynamic typing~\cite{liu2023syntax,nguyen2015divide}. Therefore, a series of learning-based code transpilers have emerged to address the limitations of traditional rule-based methods~\cite{chen2018tree,karaivanov2014phrase,nguyen2016mapping,oda2015learning}. For example, some studies adopt unsupervised or weakly supervised strategies, leveraging large-scale monolingual corpora to train models. While these approaches have achieved substantial improvements over earlier heuristic-based techniques, several challenges still remain. Specifically, one of the most critical limitation is that their translation performance still falls short of the requirements for real-world deployment. Recently, with the advent of LLMs, which demonstrate powerful generality across a wide range of code-related tasks, automated code translation has gained renewed momentum.  Recent studies reveal that LLM-based translation methods are particularly promising ~\cite{jana2024cotran,yang2024exploring,tao2024unraveling,yuan2024transagent,liu2024hmcodetrans,ibrahimzada2024alphatrans, wang2023transmap,pan2023stelocoder,luo2024bridging,pan2023understanding,luo2025unlocking,dearing2024lassi,eniser2024towards,macedo2024exploring,nitin2024spectra}. For instance, UniTrans ~\cite{yang2024exploring} improves translation accuracy by generating test cases and iteratively repairing incorrect outputs, while hmCodeTrans ~\cite{liu2024hmcodetrans} demonstrates the effectiveness of human-in-the-loop collaboration in enhancing translation quality. These methods have shown notable improvements over earlier methods, particularly in terms of execution correctness and their ability to perform end-to-end translation automatically.
    
\subsection{Balancing Accuracy and Latency in Code Translation}
    
    While most prior work on automated code translation has improved translation accuracy, LLM-based systems still struggle with complex semantic transformations. Recent approaches have integrated Chain-of-Thought (CoT) prompting strategies into the code translation task ~\cite{tao2024unraveling, yuan2024transagent}, encouraging LLMs to reason through multiple intermediate steps before producing the final output. More advanced models, such as DeepSeek-R1 ~\cite{guo2025deepseek}, are designed to generate reasoning processes without explicit prompting. Compared to DeepSeek-V3, which produces direct outputs without intermediate reasoning, DeepSeek-R1 achieves higher accuracy (from 86\% to 92\%). However, this improvement comes at the cost of a substantial 540\% increase in inference latency.
    
     These observations reveal a growing tension between reasoning-enhanced translation quality and inference efficiency. However, to the best of our knowledge, no existing framework has been specifically designed to balance the trade-off between accuracy gains and inference latency in the domain of automated code translation. 

\subsection{Internalization of CoT}

    Since the introduction of Chain-of-Thought (CoT) prompting~\cite{wei2022chain}, numerous studies have explored ways to reduce the length of reasoning traces or internalize the reasoning process altogether~\cite{kang2025c3ot,deng2024explicit,deng2023implicit,hao2024training,chen2025towards,sui2025stop,han2024token,xia2025tokenskip,ma2025cot,xu2025chain,aytes2025sketch,feng2025efficient,nayab2024concise,shen2025efficient,shen2025codi}. These efforts aim to retain the benefits of intermediate reasoning while mitigating the associated inference cost.
    
    For example, Kang et al.~\cite{kang2025c3ot} proposed C3oT, a CoT compression framework that generates shorter yet informative reasoning traces. By using a conditioned training strategy, the model learns to map long CoTs to their compressed counterparts, enabling over 50\% reduction in CoT length without compromising performance. In a more radical shift, Hao et al.~\cite{hao2024training} introduced Coconut, a latent reasoning framework that eliminates natural language CoTs altogether. Instead, the model performs reasoning directly in latent space using internal hidden states as “continuous thoughts,” which are recursively fed back into the model. This paradigm enables breadth-first exploration of reasoning paths and reduces token-level computation, showing superior performance on tasks requiring planning and backtracking. These approaches demonstrate two promising directions for reducing the cost of reasoning: (1) compressing CoT sequences in the language space, and (2) fully internalizing reasoning into latent representations. Both directions provide insights into building efficient, high-performing LLMs suitable for latency-sensitive applications.
    
    Building on these insights, our work explores an alternative internalization strategy tailored for code translation, which balances reasoning quality and inference efficiency through stepwise training.

\section{Approach}

\begin{figure*}[t]
    \centering
    \includegraphics[width=1.0\linewidth]{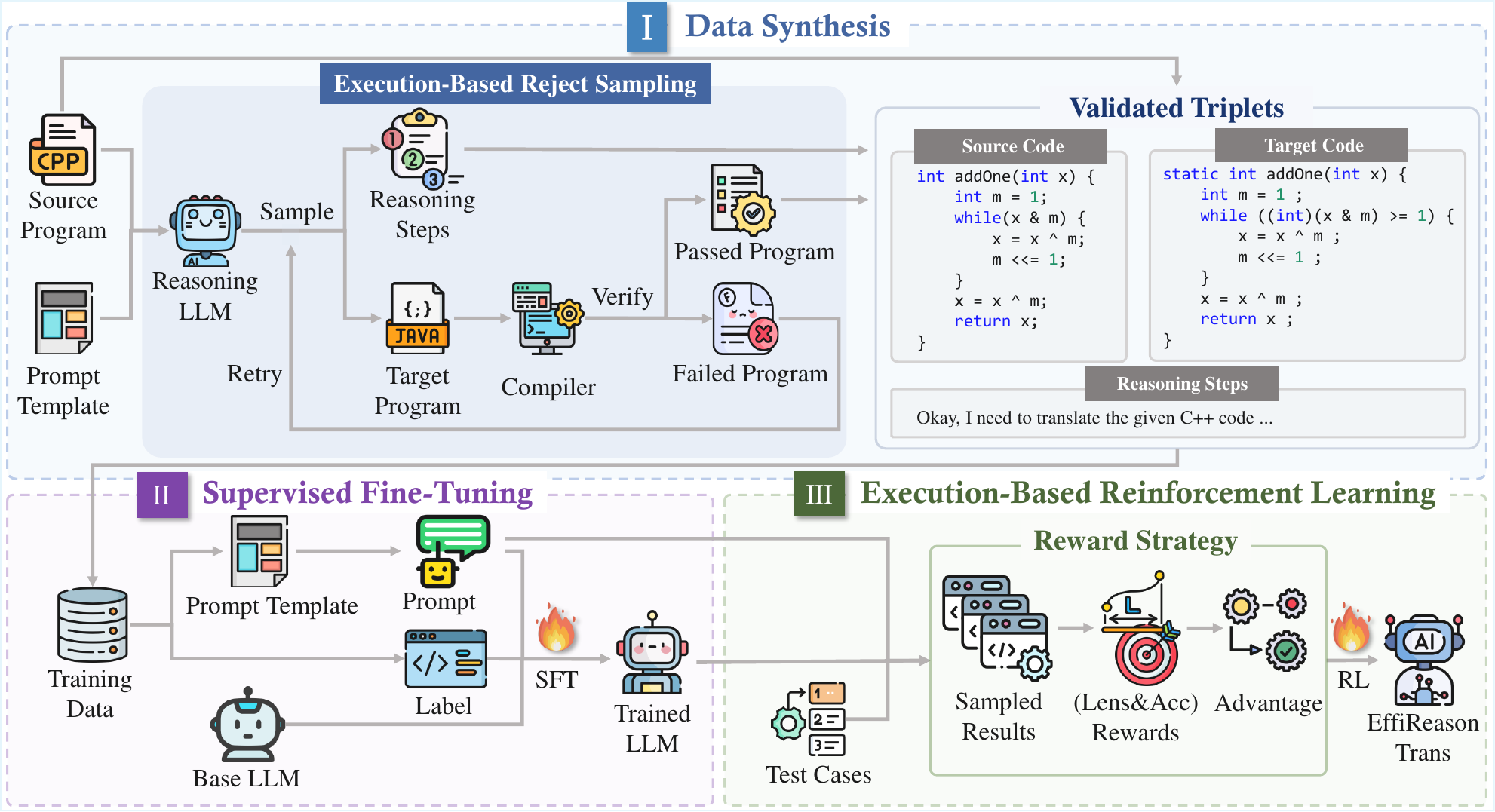}
    \vspace{-0.8em}
    \caption{Overview of \method.}
    \label{fig:overview}
    \Description{A pipeline diagram of \method.}
\end{figure*}
    
\subsection{Overview}
    In this section, we introduce \method, a training framework for code translation that aims to enhance translation accuracy while balancing inference latency. The overview of \method is shown in Figure~\ref{fig:overview}, comprising three components: a \textit{data synthesis stage}, followed by a two-stage model training process combining \textit{supervised fine-tuning} and \textit{reinforcement learning}. In the first stage, we construct a high-quality dataset by prompting a stronger LLM DeepSeek-R1~\cite{guo2025deepseek} to generate reasoning augmented translation samples. Each sample includes a triplet of source code, explicit reasoning, and target code, and is filtered through automated syntax and functionality checks to ensure reliability. Based on the synthesized dataset, called \dataset, we then perform a two-stage training process: first, the target model is fine-tuned to learn reasoning-aware code translation patterns; second, reinforcement learning is applied to further improve accuracy using reward signals derived from the model generations and the code execution results. This design aims to leverage explicit reasoning to enhance translation accuracy while mitigating the latency overhead typically introduced by longer inference chains.

\subsection{Data Synthesis}

   \begin{figure}[t]
        \centering
        \includegraphics[width=1.0\linewidth]{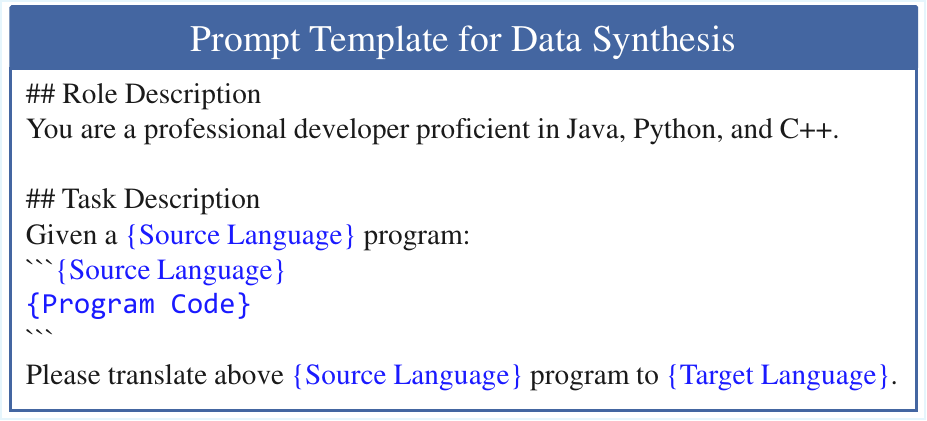}
        \vspace{-2em}
        \caption{The prompt template for data synthesis.}
        \label{fig:prompt_data_synthesis_template}
        \Description{A visual illustration of the prompt structure used during the data synthesis stage}
    \end{figure}
    
    The first component of \method is a data synthesis stage, where we construct a high-quality dataset to support reasoning-aware code translation. This stage consists of two steps: collecting clean source programs with reliable test cases, and generating reasoning augmented translation data using a reasoning-capable LLM.
    
    \subsubsection{Collecting Source Programs.}
    
    We start from a publicly available dataset hosted on Hugging Face\footnote{\url{https://huggingface.co/datasets/ziwenyd/transcoder-geeksforgeeks}}. The dataset contains parallel functions implemented in Python, Java, and C++, along with associated test cases. We perform a series of steps to filter out low-quality samples: those that fail to compile or run, or those whose tests do not pass are removed. Additionally, to prevent data leakage, we exclude samples that overlap with the test dataset used in our evaluation. After these filtering steps, we retain a curated set of 180 parallel functions across the three languages. Each function is accompanied by at least ten test cases, with average code coverage exceeding 95\% (in many cases reaching 100\%). This ensures that the test cases provide sufficient functional validation, which is crucial for later evaluation.
    
    \subsubsection{Generating Reasoning Augmented Translations.}

    To generate reasoning augmented training samples, we use the latest open-source version of DeepSeek-R1 (DeepSeek-R1-0120) \footnote{\url{https://github.com/deepseek-ai/DeepSeek-Coder}}—a reasoning-oriented large language model that extends DeepSeek-R1-Zero~\cite{guo2025deepseek}. DeepSeek-R1 introduces multi-stage training and cold-start data strategies prior to reinforcement learning, which address issues like language mixing and readability. These improvements enable the model to produce more reliable and explicit reasoning chains.
    
    For input prompts, we adopt the template as shown in Figure ~\ref{fig:prompt_data_synthesis_template}, which provides a structured translation instruction along with source code context. When queried with this prompt, DeepSeek-R1 naturally outputs two components: an explicit reasoning sequence describing the translation process, and the final translated target code.
    
    \subsubsection{Constructing Reasoning Augmented Triplets.}

    \begin{table}[t]
    \centering
    \small
        \caption{Overview of \dataset.
        }
        \label{tab:cot_dataset_stats}
        \begin{tabular}{lccc}
        \toprule
        \textbf{Translation Pair} & \textbf{\# Samples} & \textbf{\# Avg. Tokens}  \\
        \midrule
        \javatopython & 1,258 & 1311.42 \\
        \cpptojava   & 1,074 & 1217.08  \\
        \pythontocpp &   691 & 1269.12  \\
        \midrule
        \textbf{Overall} & \textbf{3,023} & \textbf{1,268.24} \\
        \bottomrule
    \end{tabular}
    \end{table}

    After validation, we construct a dataset \method $\mathcal{D} = \{(C_s, R, C_t)\}$, including 3032 training samples, where each element is a triplet composed of:
    \begin{itemize}
        \item $C_s$: The source code snippet to be translated, covering diverse data structures and algorithmic patterns. These are selected from the filtered source program set.
        \item $R$: The explicit reasoning steps produced by DeepSeek-R1 during translation generation.
        \item $C_t$: The final translated code in the target language, functionally equivalent to $C_s$, and extracted from the answer section of the model’s output.
    \end{itemize}
    \dataset offers supervision for training models to not only generate accurate translations but also to understand the rationale behind the translation process, thus laying a solid foundation for the subsequent training stages. As shown in Table~\ref{tab:cot_dataset_stats}, it comprises 1,258 \javatopython samples, 1,074 \cpptojava samples, and 691 \pythontocpp samples, covering a diverse range of translation scenarios. 
    
\subsection{Supervised Fine-Tuning}

    \begin{figure}[t]
        \centering
        \includegraphics[width=1.0\linewidth]{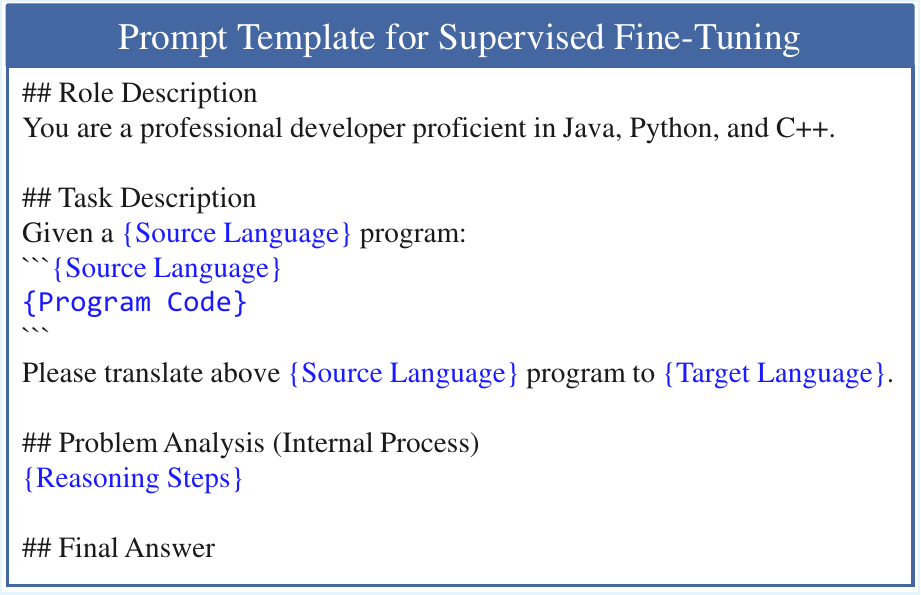}
        \vspace{-2em}
        \caption{The prompt template for supervised fine-tuning.}
        \label{fig:prompt_Sft_template}
        \Description{A visual illustration of the prompt structure used during fine-tuning, consisting of instruction, analysis, and answer segments.}
    \end{figure}
    
    In this stage, we fine-tune a reasoning-capable language model to learn how to translate source code into the target language through step-by-step reasoning. The training is conducted on the synthesized dataset \dataset, where each triplet includes the source code $C_s$, the explicit reasoning process $R$, and the translated code $C_t$.
    
    To construct each training sample, we design a structured prompt that guides the model to follow a reasoning path before producing the final output. As illustrated in Figure~\ref{fig:prompt_Sft_template}, the prompt consists of three parts: (1) task prompt: a task-specific instruction that indicates the translation direction (e.g., ``Translate the above C++ code into Java code'') and provides the input code; (2) an problem analysis section that encourages the model to analyze the semantics of the input code and plan its translation; and (3) an answer cue (e.g., “Final Answer:”) that prompts the model to begin generating the output. During supervised fine-tuning, the training labels consist solely of the translated code \( C_t \), while the chosen model, due to its inherent reasoning capability, naturally generates intermediate reasoning steps \( R \) as part of its output. Since these reasoning steps are not present in the training labels, they are treated as unconstrained outputs and are included in the decoding process without direct supervision. Despite this, we observe that the model continues to exhibit consistent reasoning behavior, which contributes to improved translation quality.

    We use DeepSeek-R1-Distill-Qwen-1.5B~\cite{guo2025deepseek} as the backbone model for fine-tuning. This model is distilled from DeepSeek-R1~\cite{guo2025deepseek}, inheriting its reasoning ability. We select the 1.5B parameter scale primarily because it significantly reduces computational and memory costs. As shown in later experiments of RQ3, this lightweight model still achieves competitive translation performance compared to models with several times more parameters.
    
    The fine-tuning is implemented using the Hugging Face Transformers library ~\cite{huggingface-docs} with the \texttt{Trainer} API. We use the standard cross-entropy loss over the entire target sequence, which includes both the reasoning steps and the translated code. To ensure proper supervision, input prompt tokens are masked during loss computation. This stage enables the model to learn reasoning-aware translation patterns in a fully supervised manner and serves as the foundation for subsequent reinforcement learning.

\subsection{Execution-Based Reinforcement Learning}

\begin{algorithm}
\caption{Execution-based Reward Strategy}
\label{alg:acc_reward}
\begin{algorithmic}[1]
\REQUIRE Completions $C$, Target Language $L$, Test Cases $T$
\ENSURE Rewards $R$
\STATE $R \leftarrow \emptyset$
\FORALL{$c$ in $C$}
    \STATE code $\leftarrow$ ExtractCode$(c, L)$
    \STATE script $\leftarrow$ PrepareTestScript$(code, T, L)$
    \STATE \textbf{try}
    \STATE \quad result $\leftarrow$ RunTestScript$(script, L)$
    \STATE \quad reward $\leftarrow \frac{\text{CountPassedTests}(result)}{\text{CountTotalTests}(result)}$
    \STATE \textbf{catch}
    \STATE \quad reward $\leftarrow 0$
    \STATE $R \leftarrow R \cup \{reward\}$
\ENDFOR
\RETURN $R$
\end{algorithmic}
\end{algorithm}

\begin{algorithm}
\caption{Length-Based Reward Strategy}
\label{alg:len_reward}
\begin{algorithmic}[1]
\REQUIRE Completions $C$, Ground Truth $G$, Max Length $M$, Tolerance $\tau$
\ENSURE Rewards $R$
\STATE $R \leftarrow \emptyset$
\FORALL{$(c, g)$ in $(C, G)$}
    \STATE $l_c \leftarrow \text{Len}(c)$, $l_g \leftarrow \text{Len}(g)$
    \IF{$l_c < l_g$ \OR $l_c > M$}
        \STATE \quad reward $\leftarrow$ 0
    \ELSIF{$\frac{l_c - l_g}{l_g} \leq \tau$}
        \STATE \quad reward $ \leftarrow 1 - \frac{l_c - l_g}{\tau \cdot l_g}$
    \ELSE
        \STATE \quad reward  $ \leftarrow$  0.1
    \ENDIF
    \STATE $R \leftarrow R \cup \{reward\}$
\ENDFOR
\RETURN $R$
\end{algorithmic}
\end{algorithm}

    In the second stage of training, we further optimize the model using reinforcement learning to enhance translation accuracy and reduce inference latency. The training data remains the same as in the previous stage, consisting of reasoning augmented triplets $\{(C_s, R, C_t)\}$that capture the translation process from source code to target code via intermediate reasoning.  Following the insights from the DeepSeek-R1~\cite{guo2025deepseek}, we initialize reinforcement learning from the supervised fine-tuned model, which ensuring stable optimization and preserves the reasoning capability acquired during SFT. Then, we adopt the GRPO algorithm~\cite{shao2024deepseekmath} for policy optimization, using the \texttt{GRPOTrainer} implementation provided by the trl library ~\cite{trl-grpo}. The input prompt format follows the same template used in the SFT stage, which showed in Figure ~\ref{fig:prompt_Sft_template}, encouraging the model to generate outputs consisting of reasoning steps followed by final translated code.
    
\subsubsection{Execution-based reward}
    To guide the policy updates, we design an execution-based reward function, shown in Algorithm~\ref{alg:acc_reward}. Specifically, after the model generates a response, we extract the translated code segment. This code is then validated against a set of test cases associated with the input function. For each sample, we calculate the reward as the fraction of test cases passed. For example, if the translated code passes 6 out of 10 test cases, the resulting reward is 0.6. This reward is used to update the model's generation policy via the GRPO algorithm. 
    
\subsubsection{Length reward}    
    To constrain the verbosity of the model outputs and encourage concise generation, we introduce a length-based auxiliary reward, as shown in Algorithm~\ref{alg:len_reward}. Given the reference solution length, we assign higher rewards to generations whose lengths fall within a relative tolerance window (e.g., ±20\% of the ground truth length). Specifically, outputs that are significantly shorter than the reference or exceed a predefined maximum length are assigned zero reward. 
    
    Overall, by incorporating reward signals derived from actual execution outcomes, the reinforcement learning stage explicitly aligns the model's generation behavior with the ultimate objective of producing semantically and functionally correct translations. In addition to the execution-based reward, we also introduce a length-based auxiliary reward to encourage concise and efficient outputs. This reward penalizes responses that are either excessively long or significantly shorter than the reference solution, while assigning higher rewards to outputs whose lengths fall within an acceptable tolerance window.

\section{EXPERIMENTS}
    In this section, we provide a detailed overview of the experimental setup. We begin by describing the computational environment, followed by an introduction to the evaluation dataset. Next, we present the baseline models used for comparison and outline the evaluation metrics employed to comprehensively assess model performance across various aspects of the code translation task.

\subsection{Experimental Setup}
\subsubsection{Computational Environment}

    For training, we use a Linux server running Ubuntu 20.04, equipped with an NVIDIA A100-SXM4-80GB GPU and CUDA 12.1. For evaluation, a separate Ubuntu 20.04 server with an NVIDIA RTX 3090 GPU (24GB VRAM) and CUDA 12.1 is used.
    
\subsubsection{Dataset}
    To assess the effectiveness of \method, we use Unitrans-Dataset\cite{yang2024exploring}. This dataset contains 568 parallel functions implemented in Python, Java, and C++, each paired with corresponding unit tests to support execution-based evaluation. In total, the dataset includes 464 unit tests for Python, 482 for Java, and 467 for C++. All samples originate from GeeksforGeeks~\cite{gfg}, a popular online platform that provides coding problems and solutions across multiple programming languages. We evaluate models across six translation pairs: \cpptopython, \cpptojava, \javatocpp, \javatopython, \pythontocpp, and \pythontojava. This diverse set of translation pairs allows us to thoroughly examine the generalizability and robustness of our method across different source-target language combinations.
    
\subsubsection{Base Model}
    As the base model, we adopt DeepSeek-R1-Distill-Qwen-1.5B~\cite{guo2025deepseek}, a distilled version of DeepSeek-R1 that retains its explicit reasoning capability while significantly reducing model size and computational cost. In our experiments, we compare the vanilla model with versions fine-tuned using our proposed framework, analyzing performance improvements across all translation pairs.

\subsubsection{Metrics}
    To comprehensively assess the performance of \method, we design dual-dimensional evaluation metrics covering \textit{effectiveness} and \textit{efficiency}. 
    
    Effectiveness focuses on how well the translated code preserves functional correctness. We assess the effectiveness using the following three metrics.
    \begin{itemize}[itemsep= 1 pt,topsep = 10 pt, parsep = 1 pt, left = 5 pt]
        \item \textbf{Computational Accuracy (CA)}: the proportion of functions whose translated code passes all associated unit tests. This and the following metric are originally proposed by Yang et al.~\cite{yang2024exploring}.
        \item \textbf{Average Pass Rate (APR)}: the average proportion of passed test cases per function, which reflects the functional correctness of translated code by evaluating how many unit tests are passed on average.
        \item \textbf{CodeBLEU}~\cite{ren2020codebleu}: an enhanced version of BLEU~\cite{papineni2002bleu} tailored for code. It incorporates syntax (via abstract syntax trees) and semantics (via data-flow analysis), and has been widely adopted in multiple studies as a standard evaluation metric for code-related tasks~\cite{lu2021codexglue,wang2021codet5}.
    \end{itemize}
    
    Efficiency is evaluated from two perspectives: the number of generated tokens and the inference latency:
    \begin{itemize} [itemsep= 1 pt,topsep = 10 pt, parsep = 1 pt, left = 5 pt]
        \item \textbf{Average Number of Generated Tokens (\#Tokens)}: We report the             average number of generated tokens per sample, denoted as:
            \[\text{\#Tokens} = \frac{1}{N} \sum_{i=1}^{N} T_i\]
            where $T_i$ is the number of tokens generated for the $i$-th input and $N$ is the total number of samples. A lower number of generated tokens typically indicates more concise output, which helps reduce memory consumption and decoding time.

        \item \textbf{Average Lnference Latency (Latency)}: We measure the average inference latency (in seconds) across the dataset, calculated as: 
                \[\text{Latency} = \frac{1}{N} \sum_{i=1}^{N} t_i\] 
             where $t_i$ is the time taken to generate a complete response for the $i$-th input. This metric directly reflects the responsiveness of the model, which is crucial for real-time applications and human-in-the-loop development scenarios.
    \end{itemize}
    Together, these metrics provide a comprehensive view of translation quality and practical deployability.

\begin{table*}[t]
\centering
\setlength\tabcolsep{10pt}
\caption{
Performance comparison across different translation pairs. Superscripts indicate relative improvements of \method over the Base model.
}
\label{tab:RQ1_translation_results}
\begin{tabular}{rclllll}
\toprule
\textbf{Translation Pair} & \textbf{Method} & \textbf{CA (\%)} & \textbf{APR (\%)} & \textbf{CodeBLEU (\%)} & \textbf{\# Tokens} & \textbf{Latency (s)} \\
\midrule
\multirow{2}{*}{\textbf{\javatopython}} 
    & Base        & 56.68 & 62.82 & 35.96 & 1389.08 & 73.40 \\
    & \cellcolor{blue!10}\textbf{\method}     
        & \cellcolor{blue!10}{\textbf{72.20}\textsuperscript{\scriptsize$\uparrow$27.4\%}} 
        & \cellcolor{blue!10}{\textbf{77.35}\textsuperscript{\scriptsize$\uparrow$23.1\%}}
        & \cellcolor{blue!10}{\textbf{39.59}\textsuperscript{\scriptsize$\uparrow$10.1\%}} 
        & \cellcolor{blue!10}{\textbf{1344.82}\textsuperscript{\scriptsize$\downarrow$3.2\%}} 
        & \cellcolor{blue!10}{\textbf{52.10}\textsuperscript{\scriptsize$\downarrow$29.0\%}} \\
\midrule
\multirow{2}{*}{\textbf{\cpptojava}} 
    & Base        & 38.17 & 40.75 & 37.90 & 1177.29 & 38.37 \\
    & \cellcolor{blue!10}\textbf{\method}     
        & \cellcolor{blue!10}{\textbf{47.30}\textsuperscript{\scriptsize$\uparrow$23.9\%}} 
        & \cellcolor{blue!10}{\textbf{49.42}\textsuperscript{\scriptsize$\uparrow$21.3\%}} 
        & \cellcolor{blue!10}{\textbf{41.01}\textsuperscript{\scriptsize$\uparrow$8.2\%}} 
        & \cellcolor{blue!10}{\textbf{1077.05}\textsuperscript{\scriptsize$\downarrow$8.5\%}} 
        & \cellcolor{blue!10}{\textbf{33.66}\textsuperscript{\scriptsize$\downarrow$12.3\%}} \\
\midrule
\multirow{2}{*}{\textbf{\pythontocpp}} 
    & Base        & 26.55 & 28.39 & 28.98 & 1494.25 & 56.33 \\
    & \cellcolor{blue!10}\textbf{\method}     
        & \cellcolor{blue!10}{\textbf{39.61}\textsuperscript{\scriptsize$\uparrow$49.2\%}} 
        & \cellcolor{blue!10}{\textbf{42.36}\textsuperscript{\scriptsize$\uparrow$49.2\%}} 
        & \cellcolor{blue!10}{\textbf{37.04}\textsuperscript{\scriptsize$\uparrow$27.8\%}} 
        & \cellcolor{blue!10}{\textbf{1205.24}\textsuperscript{\scriptsize$\downarrow$19.3\%}} 
        & \cellcolor{blue!10}{\textbf{44.02}\textsuperscript{\scriptsize$\downarrow$21.9\%}} \\
\midrule
\multirow{2}{*}{\textbf{\cpptopython}} 
    & Base        & 54.31 & 59.53 & 35.28 & 1483.81 & 56.14 \\
    & \cellcolor{blue!10}\textbf{\method}     
        & \cellcolor{blue!10}{\textbf{64.22}\textsuperscript{\scriptsize$\uparrow$18.2\%}} 
        & \cellcolor{blue!10}{\textbf{70.58}\textsuperscript{\scriptsize$\uparrow$18.6\%}} 
        & \cellcolor{blue!10}{\textbf{38.59}\textsuperscript{\scriptsize$\uparrow$9.4\%}} 
        & \cellcolor{blue!10}{\textbf{1231.41}\textsuperscript{\scriptsize$\downarrow$17.0\%}} 
        & \cellcolor{blue!10}{\textbf{42.07}\textsuperscript{\scriptsize$\downarrow$25.1\%}} \\
\midrule
\multirow{2}{*}{\textbf{\pythontojava}} 
    & Base        & 33.20 & 37.22 & 36.50 & 1347.86 & 44.57 \\
    & \cellcolor{blue!10}\textbf{\method}     
        & \cellcolor{blue!10}{\textbf{45.85}\textsuperscript{\scriptsize$\uparrow$38.1\%}} 
        & \cellcolor{blue!10}{\textbf{49.07}\textsuperscript{\scriptsize$\uparrow$31.8\%}} 
        & \cellcolor{blue!10}{\textbf{40.51}\textsuperscript{\scriptsize$\uparrow$11.0\%}} 
        & \cellcolor{blue!10}{\textbf{1199.68}\textsuperscript{\scriptsize$\downarrow$11.0\%}} 
        & \cellcolor{blue!10}{\textbf{34.89}\textsuperscript{\scriptsize$\downarrow$21.7\%}} \\
\midrule
\multirow{2}{*}{\textbf{\javatocpp}} 
    & Base        & 41.97 & 44.56 & 38.42 & 1187.64 & 46.16 \\
    & \cellcolor{blue!10}\textbf{\method}     
        & \cellcolor{blue!10}{\textbf{50.54}\textsuperscript{\scriptsize$\uparrow$20.4\%}} 
        & \cellcolor{blue!10}{\textbf{53.04}\textsuperscript{\scriptsize$\uparrow$19.0\%}} 
        & \cellcolor{blue!10}{\textbf{43.34}\textsuperscript{\scriptsize$\uparrow$12.8\%}} 
        & \cellcolor{blue!10}{\textbf{967.63}\textsuperscript{\scriptsize$\downarrow$18.5\%}} 
        & \cellcolor{blue!10}{\textbf{37.27}\textsuperscript{\scriptsize$\downarrow$19.3\%}} \\
\bottomrule
\end{tabular}
\end{table*}

\subsection{Evaluation Results}
In this section, we conduct experiments to investigate and answer the following research questions (RQs):
\begin{itemize}[itemsep= 1 pt,topsep = 10 pt, parsep = 1 pt, left = 5 pt]
\item \textbf{RQ1}:
How effective is \method in code translation?
\item \textbf{RQ2}:
How does each component impact the performance of \method?
\item \textbf{RQ3}:
Can a small model trained with \method rival the performance of a larger model?
\item \textbf{RQ4}: 
How does multilingual training data impact model performance?
\item \textbf{RQ5}: 
How does \method generalize to agent-based frameworks?
\end{itemize}

\subsubsection{\textbf{Overall Effectiveness (RQ1)}}
    \begin{figure}[t]
        \centering
        \includegraphics[width=1.0\linewidth]{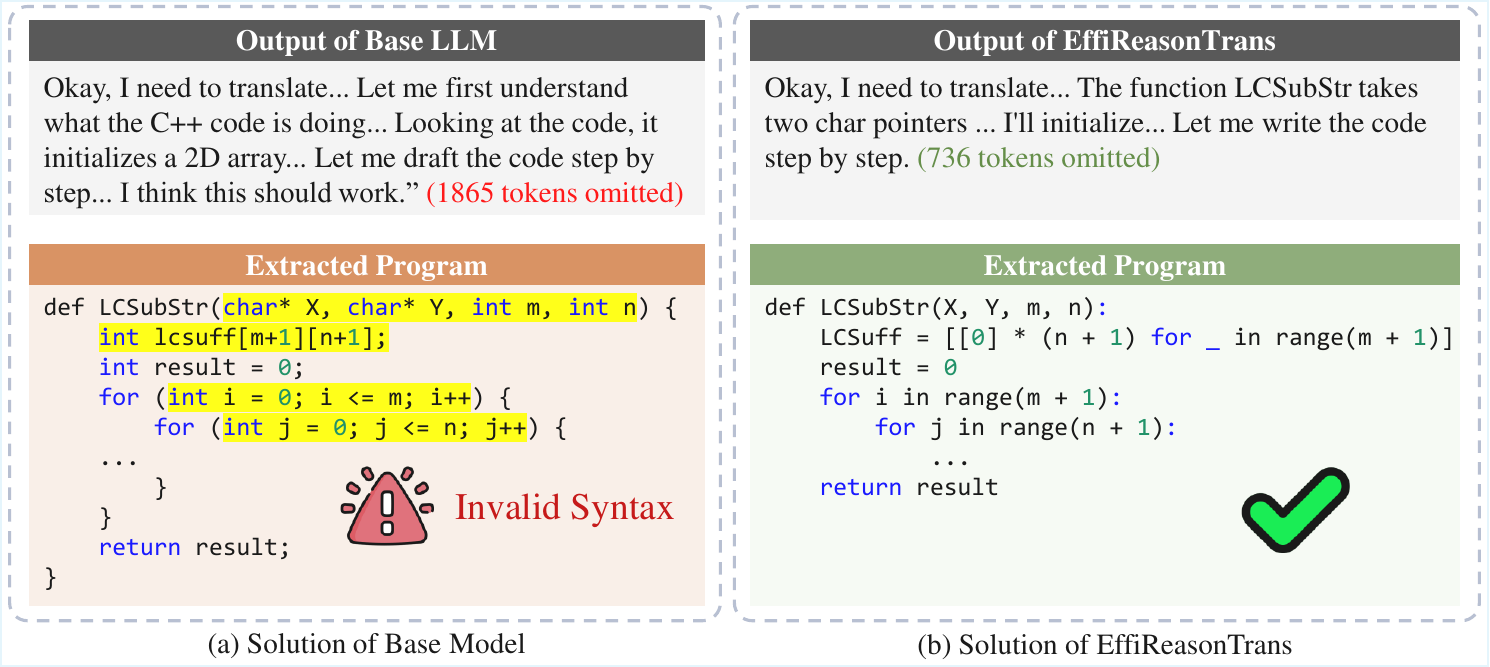}
        \vspace{-2em}
        \caption{Case study \textit{(id: LONGEST\_COMMON\_SUBSTRING)}.} 
        \label{fig:/case}
        \Description{ }
    \end{figure}
    
    To investigate whether \method can effectively improve code translation accuracy while reducing inference latency, we compare its performance with the base model (DeepSeek-R1-Distill-Qwen-1.5B) on six translation pairs using Unitrans-Dataset. We evaluate the models on both effectiveness metrics—Computational Accuracy (CA), Average Pass Rate (APR), CodeBLEU and efficiency metrics—\#Tokens and Latency.
    
    Table~\ref{tab:RQ1_translation_results} summarizes the experimental results. Across all six translation pairs, \method consistently outperforms the base model in terms of translation accuracy. Specifically, CA improvements range from 18.2\% to 49.2\%, with similar trends observed in APR (18.6\%–49.2\%) and CodeBLEU (8.2\%–27.8\%), indicating better syntactic and semantic quality of generated code.
    
    Importantly, \method achieves these accuracy improvements while substantially reducing inference latency. The average number of generated tokens decreases ranging from 3.2\% to 19.3\%, which directly contributes to lower latency. Correspondingly, average inference latency is consistently reduced (ranging from 12.3\% to 29.0\%) for all translation pairs.

    For example, Figure~\ref{fig:/case} illustrates a representative case from the \cpptopython translation task. The left side shows the output of the base model. The model directly copies C++ constructs such as “int lcsuff[m+1][n+1];” and “for (int i=0; ...)”, leading to syntax errors in Python due to mismatched language paradigms. In contrast, the right side demonstrates the output from \method, which exhibits better understanding of Python's syntax. It allocates arrays using Pythonic list comprehensions and applies correct "for i in range(...)" loops. Moreover, after applying \method, the reasoning part in this case becomes more concise (from 1865 tokens to 736 tokens). At the same time, the inference latency is reduced from 70.01 to 33.21, indicating that the simplification of reasoning contributes to lower latency. 
    
    Overall, these results demonstrate that \method successfully balances the trade-off between accuracy and efficiency. By integrating reinforcement learning with supervised fine-tuning on reasoning-augmented data, \method generates more accurate translations while producing fewer tokens, thereby reducing inference latency. This balance is critical for deploying code translation models in real-world development environments, where high accuracy and low latency are both essential.
    
    \begin{center}
        \begin{mybox}
            {\textbf{RQ1 Summary: } \method significantly improves translation accuracy (18.2\%\ \textasciitilde\ 49.2\% CA $\uparrow$) while reducing  inference latency (12.3\% \textasciitilde\ 29.0\% time $\downarrow$) across all six translation pairs, effectively balancing accuracy and efficiency.
            }
        \end{mybox}
    \end{center}

\subsubsection{\textbf{Component Contribution (RQ2)}}
\begin{table*}[t]
\centering
\setlength\tabcolsep{10pt}
\caption{
Ablation study results across different translation pairs. Superscripts indicate relative improvements over Base.
}
\label{tab:RQ2_translation_results}
\begin{tabular}{rclllll}
 \toprule
    \textbf{Translation Pair} & \textbf{Method} & \textbf{CA (\%)} & \textbf{APR (\%)} & \textbf{CodeBLEU (\%)} & \textbf{\#Tokens} & \textbf{Latency (s)} \\
    \midrule

    \multirow{4}{*}{\textbf{\javatopython}} 
        & Base      & 56.68 & 62.82 & 35.96 & 1389.08 & 73.40 \\
        & \rlonly  & {50.86\textsuperscript{\scriptsize$\downarrow$10.3\%}}
                   & {56.90\textsuperscript{\scriptsize$\downarrow$9.4\%}}
                   & {35.55\textsuperscript{\scriptsize$\downarrow$1.1\%}}
                   & {\textbf{1230.52}\textsuperscript{\scriptsize$\downarrow$11.4\%}}
                   & {72.36\textsuperscript{\scriptsize$\downarrow$1.4\%}} \\
        & \sftonly  & {71.34\textsuperscript{\scriptsize$\uparrow$25.9\%}}
                   & {76.06\textsuperscript{\scriptsize$\uparrow$21.1\%}}
                   & {39.13\textsuperscript{\scriptsize$\uparrow$8.8\%}}
                   & {1350.10\textsuperscript{\scriptsize$\downarrow$2.8\%}}
                   & {68.59\textsuperscript{\scriptsize$\downarrow$6.6\%}} \\
        & \cellcolor{blue!10}\textbf{\method}
                   & \cellcolor{blue!10}{\textbf{72.20}\textsuperscript{\scriptsize$\uparrow$27.4\%}}
                   & \cellcolor{blue!10}{\textbf{77.35}\textsuperscript{\scriptsize$\uparrow$23.1\%}}
                   & \cellcolor{blue!10}{\textbf{39.59}\textsuperscript{\scriptsize$\uparrow$10.1\%}}
                   & \cellcolor{blue!10}{1344.82\textsuperscript{\scriptsize$\downarrow$3.2\%}}
                   & \cellcolor{blue!10}{\textbf{52.10}\textsuperscript{\scriptsize$\downarrow$29.0\%}} \\
    \midrule

    \multirow{4}{*}{\textbf{\cpptojava}} 
        & Base      & 38.17 & 40.75 & 37.90 & 1177.29 & 38.37 \\
        & \rlonly  & {39.42\textsuperscript{\scriptsize$\uparrow$3.3\%}}
                   & {41.64\textsuperscript{\scriptsize$\uparrow$2.2\%}}
                   & {37.96\textsuperscript{\scriptsize$\uparrow$0.2\%}}
                   & {1194.64\textsuperscript{\scriptsize$\uparrow$1.5\%}}
                   & {70.56\textsuperscript{\scriptsize$\uparrow$83.9\%}} \\
        & \sftonly  & {45.44\textsuperscript{\scriptsize$\uparrow$19.0\%}}
                   & {46.70\textsuperscript{\scriptsize$\uparrow$14.6\%}}
                   & {41.59\textsuperscript{\scriptsize$\uparrow$9.7\%}}
                   & {\textbf{1028.18}\textsuperscript{\scriptsize$\downarrow$12.7\%}}
                   & {45.74\textsuperscript{\scriptsize$\uparrow$19.2\%}} \\
        & \cellcolor{blue!10}\textbf{\method}
                   & \cellcolor{blue!10}{\textbf{47.30}\textsuperscript{\scriptsize$\uparrow$23.9\%}}
                   & \cellcolor{blue!10}{\textbf{49.42}\textsuperscript{\scriptsize$\uparrow$21.3\%}}
                   & \cellcolor{blue!10}{\textbf{41.01}\textsuperscript{\scriptsize$\uparrow$8.2\%}}
                   & \cellcolor{blue!10}{1077.05\textsuperscript{\scriptsize$\downarrow$8.5\%}}
                   & \cellcolor{blue!10}{\textbf{33.66}\textsuperscript{\scriptsize$\downarrow$12.3\%}} \\
    \midrule

    \multirow{4}{*}{\textbf{\pythontocpp}} 
        & Base      & 26.55 & 28.39 & 28.98 & 1494.25 & 56.33 \\
        & \rlonly  & {25.48\textsuperscript{\scriptsize$\downarrow$4.0\%}}
                   & {27.17\textsuperscript{\scriptsize$\downarrow$4.3\%}}
                   & {30.89\textsuperscript{\scriptsize$\downarrow$6.6\%}}
                   & {1509.31\textsuperscript{\scriptsize$\uparrow$1.0\%}}
                   & {82.48\textsuperscript{\scriptsize$\uparrow$46.4\%}} \\
        & \sftonly  & {29.55\textsuperscript{\scriptsize$\uparrow$11.3\%}}
                   & {31.50\textsuperscript{\scriptsize$\uparrow$11.0\%}}
                   & {29.78\textsuperscript{\scriptsize$\uparrow$2.8\%}}
                   & {1397.53\textsuperscript{\scriptsize$\downarrow$6.5\%}}
                   & {45.34\textsuperscript{\scriptsize$\downarrow$19.5\%}} \\
        & \cellcolor{blue!10}\textbf{\method}
                   & \cellcolor{blue!10}{\textbf{39.61}\textsuperscript{\scriptsize$\uparrow$49.2\%}}
                   & \cellcolor{blue!10}{\textbf{42.36}\textsuperscript{\scriptsize$\uparrow$49.2\%}}
                   & \cellcolor{blue!10}{\textbf{37.04}\textsuperscript{\scriptsize$\uparrow$27.8\%}}
                   & \cellcolor{blue!10}{\textbf{1205.24}\textsuperscript{\scriptsize$\downarrow$19.3\%}}
                   & \cellcolor{blue!10}{\textbf{44.02}\textsuperscript{\scriptsize$\downarrow$21.9\%}} \\
    \midrule
    \multirow{4}{*}{\textbf{\cpptopython}} 
        & Base      & 54.31 & 59.53 & 35.28 & 1483.81 & 56.14 \\
        & \rlonly   & {49.14\textsuperscript{\scriptsize$\downarrow$9.5\%}}
                    & {55.56\textsuperscript{\scriptsize$\downarrow$6.7\%}}
                    & {34.61\textsuperscript{\scriptsize$\downarrow$1.9\%}}
                    & {1499.25\textsuperscript{\scriptsize$\uparrow$1.0\%}}
                    & {81.95\textsuperscript{\scriptsize$\uparrow$46.0\%}} \\
        & \sftonly  & {63.15\textsuperscript{\scriptsize$\uparrow$16.3\%}}
                    & {68.23\textsuperscript{\scriptsize$\uparrow$14.6\%}}
                    & {37.91\textsuperscript{\scriptsize$\uparrow$7.5\%}}
                    & {1265.65\textsuperscript{\scriptsize$\downarrow$14.7\%}}
                    & {44.40\textsuperscript{\scriptsize$\downarrow$20.9\%}} \\
        & \cellcolor{blue!10}\textbf{\method}
                    & \cellcolor{blue!10}{\textbf{64.22}\textsuperscript{\scriptsize$\uparrow$8.2\%}}
                    & \cellcolor{blue!10}{\textbf{70.58}\textsuperscript{\scriptsize$\uparrow$18.6\%}}
                    & \cellcolor{blue!10}{\textbf{38.59}\textsuperscript{\scriptsize$\uparrow$9.4\%}}
                    & \cellcolor{blue!10}{\textbf{1231.41}\textsuperscript{\scriptsize$\downarrow$17.0\%}}
                    & \cellcolor{blue!10}{\textbf{42.07}\textsuperscript{\scriptsize$\downarrow$25.1\%}} \\
    \midrule

    \multirow{4}{*}{\textbf{\pythontojava}} 
        & Base      & 33.20 & 37.22 & 36.50 & 1347.86 & 44.57 \\
        & \rlonly   & {35.89\textsuperscript{\scriptsize$\uparrow$8.1\%}}
                    & {40.21\textsuperscript{\scriptsize$\uparrow$8.0\%}}
                    & {36.75\textsuperscript{\scriptsize$\uparrow$0.7\%}}
                    & {1363.22\textsuperscript{\scriptsize$\uparrow$1.1\%}}
                    & {76.99\textsuperscript{\scriptsize$\uparrow$72.7\%}} \\
        & \sftonly  & {42.74\textsuperscript{\scriptsize$\uparrow$28.7\%}}
                    & {46.41\textsuperscript{\scriptsize$\uparrow$24.7\%}}
                    & {40.08\textsuperscript{\scriptsize$\uparrow$9.8\%}}
                    & {\textbf{1083.93}\textsuperscript{\scriptsize$\downarrow$19.6\%}}
                    & {38.21\textsuperscript{\scriptsize$\downarrow$14.3\%}} \\
        & \cellcolor{blue!10}\textbf{\method}
                    & \cellcolor{blue!10}{\textbf{45.85}\textsuperscript{\scriptsize$\uparrow$38.1\%}}
                    & \cellcolor{blue!10}{\textbf{49.07}\textsuperscript{\scriptsize$\uparrow$31.8\%}}
                    & \cellcolor{blue!10}{\textbf{40.51}\textsuperscript{\scriptsize$\uparrow$11.0\%}}
                    & \cellcolor{blue!10}{1199.68\textsuperscript{\scriptsize$\downarrow$11.0\%}}
                    & \cellcolor{blue!10}{\textbf{34.89}\textsuperscript{\scriptsize$\downarrow$21.7\%}} \\
    \midrule

    \multirow{4}{*}{\textbf{\javatocpp}} 
        & Base      & 41.97 & 44.56 & 38.42 & 1187.64 & 46.16 \\
        & \rlonly   & {41.97\textsuperscript{\scriptsize$\pm$0.0\%}}
                    & {45.18\textsuperscript{\scriptsize$\uparrow$1.4\%}}
                    & {37.36\textsuperscript{\scriptsize$\downarrow$2.8\%}}
                    & {1230.71\textsuperscript{\scriptsize$\uparrow$3.6\%}}
                    & {72.36\textsuperscript{\scriptsize$\uparrow$56.8\%}} \\
        & \sftonly  & {44.97\textsuperscript{\scriptsize$\uparrow$7.1\%}}
                    & {48.20\textsuperscript{\scriptsize$\uparrow$8.2\%}}
                    & {42.16\textsuperscript{\scriptsize$\uparrow$9.7\%}}
                    & {985.86\textsuperscript{\scriptsize$\downarrow$17.0\%}}
                    & {49.95\textsuperscript{\scriptsize$\uparrow$8.2\%}} \\
        & \cellcolor{blue!10}\textbf{\method}
                    & \cellcolor{blue!10}{\textbf{50.54}\textsuperscript{\scriptsize$\uparrow$20.4\%}}
                    & \cellcolor{blue!10}{\textbf{53.04}\textsuperscript{\scriptsize$\uparrow$19.0\%}}
                    & \cellcolor{blue!10}{\textbf{43.34}\textsuperscript{\scriptsize$\uparrow$12.8\%}}
                    & \cellcolor{blue!10}{\textbf{967.63}\textsuperscript{\scriptsize$\downarrow$18.5\%}}
                    & \cellcolor{blue!10}{\textbf{37.27}\textsuperscript{\scriptsize$\downarrow$19.3\%}} \\
    \bottomrule
\end{tabular}
\end{table*}

To investigate how the main components (supervised fine-tuning and reinforcement learning) of \method contribute to its performance, we conduct ablation studies by comparing the following four training settings:
\begin{itemize}[itemsep= 1 pt,topsep = 10 pt, parsep = 1 pt, left = 5 pt]
    \item \textbf{Base}: The pretrained base model (DeepSeek-R1-Distill-1.5B) without fine-tuning.
    \item \textbf{\sftonly}: Base model with only supervised fine-tuning on reasoning-augmented data.
    \item \textbf{\rlonly}: Base model with only reinforcement learning (no SFT).
    \item \textbf{\method}: Our full two-stage approach (SFT followed by RL).
\end{itemize}
    
All the three training strategies are implemented on the synthesized reasoning augmented dataset \dataset.
Based on the experimental results shown in Table~\ref{tab:RQ2_translation_results}, we have the following observations:

    \textbf{(1) Supervised fine-tuning lays a strong foundation} . Across all translation pairs, supervised fine-tuning leads to noticeable improvement over the base setting and \rlonly. For example, in the \javatopython task, \sftonly improves CA from 50.86\% (Base) and 56.68\% (\rlonly) to 71.34\%, achieving relative gains of 40.0\% and 25.9\ respectively. Similar trends are observed in other pairs. Compared to the base setting, \sftonly also help to decrease the number of generated tokens and inference latency in most cases, such as \pythontocpp (\#Tokens$\downarrow$ 6.5\%, Latency$\downarrow$ 19.5\%).

    \textbf{(2) Reinforcement learning directly shows limited improvement in accuracy}.From the experimental results, we observe that directly applying reinforcement shows limited imporvement in accuracy. For some translation pairs, such as \javatopython, \rlonly even decrease from 56.68\% to 50.86\% in CA. We speculate that this phenomenon can be attributed to the absence of a proper supervised fine-tuning (SFT) initialization, causing the reinforcement learning process to start from a relatively random or weak policy. This leads to a large exploration space and unstable gradient updates, which often result in the model converging to suboptimal policies or local minima. Therefore, incorporating an SFT stage prior to RL provides a stronger initialization, ultimately improving both the overall accuracy and the convergence speed.

    \textbf{(3) \method offers improved effectiveness-efficiency trade-offs}.
    The full two-stage approach generally achieves better performance by improving accuracy significantlt while generally reducing the number of generated tokens and inference latency. For example, in the \pythontocpp task, \method achieves significant improvements from 26.55\% to 39.61\% (+49.2\%) in CA while decreases Latency from 56.33s to 44.02s (-21.9\%). Similar trends are observed across other translation pairs.

    \begin{center}
        \begin{mybox}
            {\textbf{RQ2 Summary: } 
            Directly applying reinforcement learning to the base model yields limited improvements. In contrast, supervised fine-tuning plays a crucial role in the two-stage training process. After supervised fine-tuning, reinforcement learning further refines and enhances the model's behavior.
            }
        \end{mybox}
    \end{center}

\subsubsection{\textbf{Performance Comparability Between Small-Scale And Larger-Scale Models (RQ3)}}

    \begin{figure*}[t]
        \centering
        \includegraphics[width=1.0\linewidth]{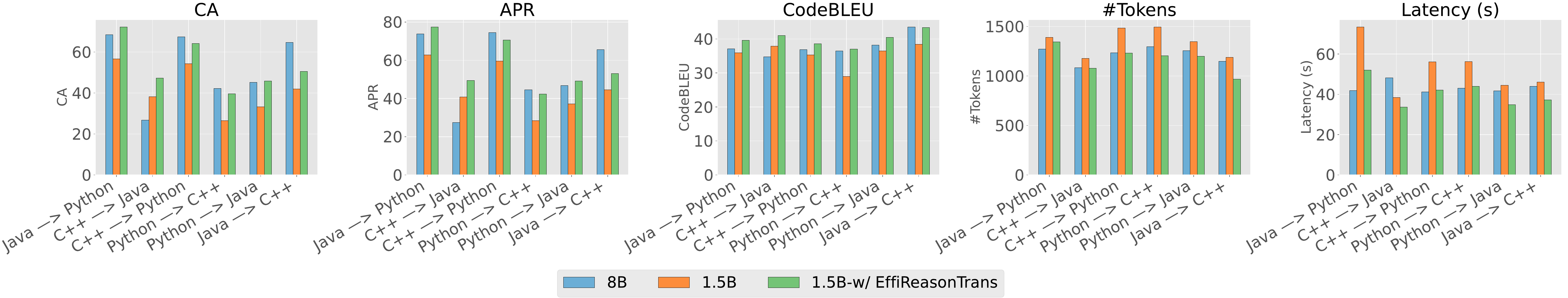}
        \vspace{-2em}
        \caption{Performance of EffiReasonTrans-enhanced 1.5B model vs. 8B model across six translation pairs.}
        \label{fig:/RQ3}
        \Description{ }
    \end{figure*}

    To investigate whether a small-scale model equipped with \method can achieve performance comparable to a larger-scale model, we compare two models of different sizes: the 8B-parameter DeepSeek-R1-Distill-Llama-8B and the 1.5B-parameter DeepSeek-R1-Distill-Qwen-1.5B~\cite{guo2025deepseek}. According to the results in DeepSeek-R1 report~\cite{guo2025deepseek}, DeepSeek-R1-Distill-Llama-8B consistently outperforms DeepSeek-R1-Distill-Qwen-1.5B across several benchmarks, including AIME 2024~\cite{aime2024}, MATH-500~\cite{guo2025deepseek}, GPQA Diamond\cite{rein2023gpqa}, LiveCode~\cite{jain2024livecodebench}, CodeForces~\cite{codeforces}, LiveCodeBench~\cite{jain2024livecodebench}.

    As shown in Figure~\ref{fig:/RQ3}, our \method-enhanced 1.5B model (denoted as ``1.5B-\method'') consistently narrows the performance gap with the 8B model across all translation pairs and metrics. Specifically, in the \javatopython task, the 1.5B-\method model achieves a CA of 72.20\%, surpassing the 8B model’s 68.53\%, along with higher APR and CodeBLEU scores. A similar trend is observed in the \cpptojava task and the \pythontojava task, where the smaller model also outperforms the 8B counterpart. Moreover, the 1.5B-\method model leads to notable performance gains over the 1.5B-Base model in the remaining three translation tasks, significantly reducing the gap with the 8B model. These improvements are often accompanied by reductions in either the average number of output tokens or inference latency, demonstrating enhanced efficiency in addition to improved translation quality.

    \begin{center}
        \begin{mybox}
            {\textbf{RQ3 Summary: } 
               \method enable small models (1.5B) to achieve comparable or superior performance to larger models (8B), with \smallwithmethod outperforming the 8B model in \javatopython and \cpptojava tasks while reducing latency by 3.2$\times$. This demonstrates that reasoning optimazation can effectively compensate for model scale reduction, making high-accuracy translation feasible on resource-constrained devices. 
            }
        \end{mybox}
    \end{center}

\subsubsection{\textbf{Multilingual Training Impact (RQ4)}} 
    \begin{figure*}[t]
        \centering
        \includegraphics[width=1.0\linewidth]{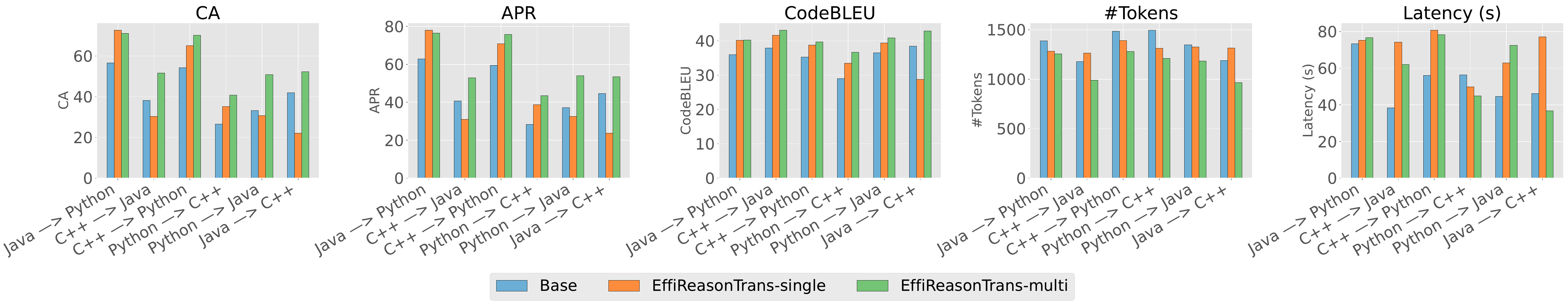}
        \vspace{-2em}
        \caption{Performance comparison of different compositions of training data.}
        \label{fig:/RQ4}
        \Description{ }
    \end{figure*}
    
    To investigate the impact of multilingual training data, we compare three variants of our method:
        \begin{itemize}[itemsep= 1 pt,topsep = 10 pt, parsep = 1 pt, left = 5 pt]
        \item \textbf{Base}: the original base model (DeepSeek-R1-Distill-1.5B) without any fine-tuning.

        \item \textbf{\method-single}: the model trained using our proposed \method on single-translation-pair data. Specifically, we collect 1,089 samples for the Java-to-Python translation pair based on the data synthesis procedure described earlier.
        
        \item \textbf{\method-multi}: the model trained using \method on multilingual data covering multiple translation pairs. We collect 1,089 samples in total, including 603 for Java-to-Python, 603 for C++-to-Java, and 603 for Python-to-C++, all constructed using the same data synthesis approach.
        \end{itemize}

    The experimental results are presented in Figure~\ref{fig:/RQ4}. We analyze the outcomes from two perspectives, as detailed below.
    
    In terms of effectiveness, \method-single exhibits large performance variance across tasks. For instance, \method-single performs well in \javatopython but poorly in \pythontojava (CA drops to 30.71\%, even lower than the base model's 33.20\%).  In contrast, \method-multi maintains strong performance across all tasks. For instance, in the \pythontojava task, \method-multi achieves 50.83\% CA, outperforming both the base model (33.20\% CA) and \method-single (30.71\% CA). This suggests that training on a single translation pair lacks generalization ability, while \method-multi benefits from cross-lingual knowledge and shows robust gains across diverse translation pairs.

    In terms of efficiency, \method-multi reduces the number of generated tokens and inference latency compared to the base model and \method-single in some cases. For example, in the \javatocpp task, \method-multi reduces average tokens from 1187.64 (base) to 966.20, and latency from 46.16s to 36.74s. Similarly, in \pythontocpp, both token length and latency are significantly reduced (1494.25 → 1209.63 tokens; 56.33s → 44.92s).

    \begin{center}
        \begin{mybox}
            {\textbf{RQ4 Summary: } 
                In summary, training on multilingual data improves both the effectiveness and generalization of the model across diverse translation tasks, while also offering potential gains in inference efficiency.
                
            }
        \end{mybox}
    \end{center}

\subsubsection{\textbf{Generalization to Agent-based Framework (RQ5)}} 

    \begin{table*}[t]
        \centering
      \setlength\tabcolsep{10pt}
        \caption{
            Performance of base and \method across different translation pairs in the agent-based setting. 
        Relative changes of \method are shown as superscript arrows to the right of values.
        }
        \label{tab: RQ5_agent_results_round2}
        \begin{tabular}{rclllll}
        \toprule
        \textbf{Translation pair} & \textbf{Method} & \textbf{CA (\%)} & \textbf{APR (\%)} & \textbf{CodeBLEU (\%)} & \textbf{\#Tokens} & \textbf{Latency (s)} \\
        \midrule
        
        \multirow{2}{*}{\textbf{\javatopython}} 
            & Base        & 55.60 & 63.34 & 34.58 & \textbf{973.72}  & \textbf{36.96} \\
            & \cellcolor{blue!10}\textbf{\method}     
                & \cellcolor{blue!10}{\textbf{73.49}\textsuperscript{\scriptsize$\uparrow$32.2\%}} 
                & \cellcolor{blue!10}{\textbf{79.98}\textsuperscript{\scriptsize$\uparrow$26.3\%}} 
                & \cellcolor{blue!10}{\textbf{39.82}\textsuperscript{\scriptsize$\uparrow$15.2\%}} 
                & \cellcolor{blue!10}{1235.12\textsuperscript{\scriptsize$\uparrow$26.8\%}} 
                & \cellcolor{blue!10}{69.98\textsuperscript{\scriptsize$\uparrow$89.3\%}} \\
        \midrule    
        \multirow{2}{*}{\textbf{\cpptojava}} 
            & Base        & 42.36  & 44.73  & 36.52 & \textbf{1030.29} & \textbf{46.31} \\
            & \cellcolor{blue!10}\textbf{\method}     
                & \cellcolor{blue!10}{\textbf{45.44}\textsuperscript{\scriptsize$\uparrow$6.8\%}} 
                & \cellcolor{blue!10}{\textbf{48.05}\textsuperscript{\scriptsize$\uparrow$6.9\%}} 
                & \cellcolor{blue!10}{\textbf{40.77}\textsuperscript{\scriptsize$\uparrow$10.4\%}} 
                & \cellcolor{blue!10}{1464.05\textsuperscript{\scriptsize$\uparrow$42.1\%}} 
                & \cellcolor{blue!10}{72.00\textsuperscript{\scriptsize$\uparrow$55.4\%}} \\
        \midrule    
        \multirow{2}{*}{\textbf{\pythontocpp}} 
            & Base        & 21.20 & 23.40 & 28.19 & \textbf{1050.44} & 51.69 \\
            & \cellcolor{blue!10}\textbf{\method}     
                & \cellcolor{blue!10}{\textbf{27.84}\textsuperscript{\scriptsize$\uparrow$31.3\%}} 
                & \cellcolor{blue!10}{\textbf{30.02}\textsuperscript{\scriptsize$\uparrow$28.3\%}} 
                & \cellcolor{blue!10}{\textbf{32.22}\textsuperscript{\scriptsize$\uparrow$14.3\%}} 
                & \cellcolor{blue!10}{1360.57\textsuperscript{\scriptsize$\uparrow$29.5\%}} 
                & \cellcolor{blue!10}{\textbf{33.22}\textsuperscript{\scriptsize$\downarrow$35.7\%}} \\
        \midrule    
        \multirow{2}{*}{\textbf{\cpptopython}} 
            & Base        & 54.09 & 61.12 & 34.00 & \textbf{1089.06} & \textbf{50.59} \\
            & \cellcolor{blue!10}\textbf{\method}     
                & \cellcolor{blue!10}{\textbf{72.63}\textsuperscript{\scriptsize$\uparrow$34.3\%}} 
                & \cellcolor{blue!10}{\textbf{79.07}\textsuperscript{\scriptsize$\uparrow$29.4\%}} 
                & \cellcolor{blue!10}{\textbf{39.27}\textsuperscript{\scriptsize$\uparrow$15.5\%}} 
                & \cellcolor{blue!10}{1112.02\textsuperscript{\scriptsize$\uparrow$2.1\%}} 
                & \cellcolor{blue!10}{68.81\textsuperscript{\scriptsize$\uparrow$36.0\%}} \\
        \midrule    
        \multirow{2}{*}{\textbf{\pythontojava}} 
            & Base        & 36.50 & 39.00 & 35.70 & \textbf{1300.00} & \textbf{48.00} \\
            & \cellcolor{blue!10}\textbf{\method}     
                & \cellcolor{blue!10}{\textbf{45.23}\textsuperscript{\scriptsize$\uparrow$23.9\%}} 
                & \cellcolor{blue!10}{\textbf{47.82}\textsuperscript{\scriptsize$\uparrow$22.6\%}} 
                & \cellcolor{blue!10}{\textbf{39.82}\textsuperscript{\scriptsize$\uparrow$11.5\%}} 
                & \cellcolor{blue!10}{1458.43\textsuperscript{\scriptsize$\uparrow$12.2\%}} 
                & \cellcolor{blue!10}{61.67\textsuperscript{\scriptsize$\uparrow$28.5\%}} \\
         \midrule   
        \multirow{2}{*}{\textbf{\javatocpp}} 
            & Base        & 36.53 & 38.71 & 34.53 & 1074.42 & \textbf{40.80} \\
            & \cellcolor{blue!10}\textbf{\method}     
                & \cellcolor{blue!10}{\textbf{40.04}\textsuperscript{\scriptsize$\uparrow$9.6\%}} 
                & \cellcolor{blue!10}{\textbf{41.88}\textsuperscript{\scriptsize$\uparrow$8.2\%}} 
                & \cellcolor{blue!10}{\textbf{36.67}\textsuperscript{\scriptsize$\uparrow$6.2\%}} 
                & \cellcolor{blue!10}{\textbf{1069.00}\textsuperscript{\scriptsize$\downarrow$0.5\%}} 
                & \cellcolor{blue!10}{62.52\textsuperscript{\scriptsize$\uparrow$53.2\%}} \\
        
        \bottomrule
        \end{tabular}
    \end{table*}

    Recent advances in code translation have introduced agent-based frameworks that leverage LLMs to iteratively refine translations based on execution feedback~\cite{yang2024exploring, ibrahimzada2024alphatrans}. In this subsection, we adopt the UniTrans framework~\cite{yang2024exploring} to assess the effectiveness of \method in the agent-based setting. UniTrans generates a set of test cases using LLM and incorporates these test cases into a prompt that instructs the LLM to translate the source code into target code aiming to pass the given test cases. Subsequently, through execution feedback, the LLM is guided to fix errors in the translated code. Referring to the original configuration, we generate three test cases per example and conduct experiments over two interaction rounds on six translation pairs.
    
    Table~\ref{tab: RQ5_agent_results_round2}  shows the final results of Unitrans, comparing the base model (DeepSeek-R1-Distill-Qwen-1.5B) and our \method-enhanced model. Based on the experimental results, we summarize the following observations:
    
    \textbf{(1) Substantial improvements in execution correctness.}
    \method consistently improves execution-based metrics (CA and APR) across nearly all translation pairs. For instance, in the \javatopython task, CA increases from 55.6\% to 73.49\% (+32.2\%), and APR improves from 63.34\% to 79.98\% (+26.3\%). These results demonstrate that \method enables the model to generate more functionally correct programs, which is critical in agent-based frameworks where the overall performance largely depends on the deployed LLM.

   \textbf{(2) Mixed impact on efficiency-related metrics.}
    While \method brings substantial improvements in execution correctness, its performance on efficiency metrics such as \#Tokens and Latency is less consistent. Specifically, in the \javatocpp task, \method achieve smaller number of generated tokens than the base model (1074.42 to 1069.00 \#Tokens) and gains improvements of CA (36.53\% to 40.04\%), indicating that the model produces more concise and effective translations when guided by the agent framework. However, despite the reduced token length, the Latency is still notably higher than the base, possibly due to more computationally intensive reasoning steps induced by \method. What's more, both \#Tokens and Latency show a noticeable increase compared to the base model across most translation pairs. This degradation suggests that the additional round of interaction amplifies the computational burden introduced by \method, possibly because the model generates more elaborate fixes in response to failed test cases or overfits to verbose reasoning patterns. 
    
    Overall, these observations reveal that while \method improves functional correctness, it may introduce an inference-time cost, particularly in multi-round agent-based workflows. This trade-off highlights the challenge of balancing accuracy and efficiency in practical deployments.

    \begin{center}
        \begin{mybox}
            {\textbf{RQ5 Summary: } 
                \method improves execution correctness in agent-based frameworks, consistently enhancing Computational Accuracy and Average Pass Rate across translation pairs and rounds. However, it also increases inference latency and output length, particularly in later rounds, revealing a trade-off between accuracy and efficiency in multi-round workflows.
            }
        \end{mybox}
    \end{center}

\section{Threats to Validity}
    Despite our efforts to conduct a comprehensive evaluation, several factors may pose threats to the validity of our findings, including the reliability of synthesized data, model selection, and programming language coverage.

    One potential threat lies in  the reliance on reasoning-augmented data automatically generated during the data synthesis stage. \method relies on automatically generated reasoning-augmented data in the data synthsis stage to train models. We utilize a stronger language model to generaete output with reasoning and veify them based on exection, which aiming to filter the invalid outputs and make sure the reasoning-augmented data more realiable. However, the generated reasoning process may still introduce hallucinated steps and guaranteeing the correctness of all reasoning steps remains challenging, due to the large amoumt of text information. Future work could incorporate stronger verficiation method to furthur detect the deeper reliability of the detailed reasoning.
        
    Another potential treate to the validity is that all experiments are conducted using a single base model, DeepSeek-R1-Distill-Qwen-1.5B. We choose the model for two main reasons. First, it demonstrates reasonable reasoning ability without additional pretraining, making it a meaningful and comparable baseline for evaluating our training framework. Second, experiments involving reinforcement learning and supervised fine-tuning are computationally expensive; focusing on a single lightweight yet capable model allowed us to conduct comprehensive evaluations across multiple translation pairs and perform detailed ablation studies within our available resources.  While our current experiments are based on a single model, we acknowledge that a more comprehensive evaluation of our method requires testing it on diverse model architectures. Therefore, we expect future work to explore this direction to more fully assess the effectiveness and generalizability of \method.

    Finally, a potential threat is the choice of programming languages used for evaluation. In this work, we focus on three widely-used languages: Python, Java, and C++. These languages are selected due to their high popularity in both academic and industrial settings and facilitates comparison with prior work. However, we acknowledge that this choice may limit the generalizability of our conclusions to other languages, particularly those low resource programing languages (e.g., ArkTS). Extending our evaluation to a broader set of programming languages is a promising direction for future work and would further verify the robustness of our method in more diverse translation scenarios.

    In summary, while we have addressed several potential concerns, some aspects such as data quality, reliance on a single base model, and the range of programming languages evaluated remain areas for further exploration. These factors highlight important avenues for future research to enhance the robustness, generalizability, and practical applicability of \method. We encourage further investigations incorporating stronger verification techniques, diverse model architectures, and a wider variety of programming languages to build upon our findings and drive advancements in reasoning-enhanced code translation.

\section{CONCLUSION}

In this paper, we propose \method, a reasoning-enhanced training framework that balances accuracy and efficiency in code translation. It comprises three stages: reasoning-augmented data synthesis, supervised fine-tuning, and reinforcement learning. High-quality (source code, reasoning, target code) triplets are generated by a powerful LLM and filtered via syntax and functional tests. Moreover, a dual-objective reward strategy is introduced to optimize both execution correctness and output conciseness. Experiments on six translation pairs show that \method consistently improves accuracy while generally reducing inference latency. Ablation and extension studies further highlight the contributions of each component and demonstrate effectiveness in multilingual and agent-based settings, suggesting \method’s practical value for real-world development workflows.

\bibliographystyle{ACM-Reference-Format}
\bibliography{ref}

\end{document}